\documentclass[12pt,preprint,notoc,nohyper]{AKRSX} % 10pt is ignored!

\usepackage{epsfig,multicol,bbm}

\newcommand\fverb{\setbox\pippobox=\hbox\bgroup\verb}
\newcommand\fverbdo{\egroup\medskip\noindent%
            \fbox{\unhbox\pippobox}\ }
\newcommand\fverbit{\egroup\item[\fbox{\unhbox\pippobox}]}
\newbox\pippobox
\title{Exact differential and quantum corrections of entropy for axially symmetric black holes}

\author{M. Akbar$^1$~, K. Saifullah$^2$ \\

$^1$Centre for Advanced Mathematics and Physics \\
National University of Sciences and Technology, Rawalpindi,
Pakistan \\

$^2$Department of Mathematics, Quaid-i-Azam University, Islamabad,
Pakistan\\

Electronic address: \email{makbar@camp.nust.edu.pk},
\email{saifullah@qau.edu.pk}}

\preprint{}  % OR: \preprint{Aaaa/Mm/Yy\\Aaa-aa/Nnnnnn}
\abstract{Using the exactness criteria of entropy from the first law
of black hole thermodynamics, we study quantum corrections for
axially symmetric black holes.}

\dedicated{}

\begin{document}

Hawking's work on black hole radiation and evaporation proved
extremely significant for studying black holes as thermodynamical
systems. Here we study quantum mechanical phenomenon in the context
of classical theory of general relativity. We are interested in
studying the changes in classical entropy of black hole due to these
quantum effects. For a black hole of mass, $M$, angular momentum,
$J$, and charge, $Q$, the first law of thermodynamics is
$dM=TdS+\Omega dJ+\Phi dQ$, where $T$ is the temperature, $S$
entropy, $\Omega$ angular velocity and $\Phi$ electrostatic
potential. We can also write this as
\begin{equation}\label{2laws}
   dS(M, J, Q)=\frac{1}{T}dM-\frac{\Omega}{T}dJ-\frac{\Phi}{T}dQ .
\end{equation}
Now,  we note that this differential in three parameters is exact if
the following conditions are satisfied \cite{MAKS-EPJC}

\begin{eqnarray}\label{3conda}
    \frac{\partial }{\partial J}\left(\frac{1}{T}\right)
=\frac{\partial }{\partial M} \left(-\frac{\Omega}{T}\right),  \\
\label{3condb} \frac{\partial }{\partial
Q}\left(\frac{1}{T}\right)=\frac{\partial }{\partial M}
\left(-\frac{\Phi}{T}\right),
\\ \label{3condc} \frac{\partial}{\partial
Q}\left(-\frac{\Omega}{T}\right)=\frac{\partial }{\partial
J}\left(-\frac{\Phi}{T}\right) .
\end{eqnarray}
Thus entropy $S(M,J,Q)$ can be written in the integral form. Using
this we work out quantum corrections of entropy \cite{solod1,
Banerjee08} beyond the semiclassical limit. Here we will apply this
to axially symmetric static spacetimes.

%\section{The Kerr-Newman black hole}

We first consider the Kerr-Newman spacetime in Boyer-Lindquist
coordinates

\begin{eqnarray*}
ds^{2} &=& -\frac{\Delta^2}{\rho^2}(dt-asin^2 \theta
d\phi)^2+\frac{\rho^2}{\Delta^2}dr^2+\rho^2d\theta^{2}+
\frac{sin^2\theta}{\rho^2}(adt-(r^2+a^2)d\phi)^2 ,
\end{eqnarray*}
where $\Delta^2 = (r^2+a^2)-2Mr+Q^2$, $\rho^2 = r^2 +
a^2cos^2\theta$ and $a = \frac{J}{M}$.

The horizons for this metric are $r_\pm=M\pm\sqrt{M^2-a^2-Q^2}$. The
outer horizon at $r_{+}$ is specified as the black hole horizon and
is a null stationary 2-surface. The Killing vector normal to this
surface is $\chi^{\alpha} = t^{\alpha} + \Omega \phi^{\alpha}$ and
it is null on the horizon. This horizon is generated by the Killing
vector $\chi^{\alpha}$, and the surface gravity $\kappa$ associated
with this Killing horizon is $\kappa^{2} = \frac{-1}{2}
\chi^{\alpha;\beta}\chi_{\alpha;\beta}$. Using this it is easy to
evaluate the temperature $T = \kappa/2$ associated with this horizon
as \cite{MAKS-EPJC}
\begin{equation}\label{temp}
 T= \left(\frac{\hbar}{2\pi}\right) \frac{\sqrt{M^4-J^2-Q^2 M^2 }}{M\left(2M^2-Q^2+
 2\sqrt{M^4-J^2-Q^2 M^2}\right)} .
\end{equation}
The angular velocity is $\Omega=
J/M\left(2M^2-Q^2+2\sqrt{M^4-J^2-Q^2 M^2 }\right)$, and the
electrostatic potential becomes

\begin{equation}\label{phi}
 \Phi= \frac{Q \left(M^2+\sqrt{M^4-J^2-Q^2 M^2 }\right)}
{M\left(2M^2-Q^2+2\sqrt{M^4-J^2-Q^2 M^2 }\right)} .
\end{equation}

It is easy to see that these quantities for the Kerr-Newman black
hole satisfy conditions (\ref{3conda})-(\ref{3condc}), and
therefore, $dS$ is an exact differential. We use the modified
surface gravity \cite{York85} due to quantum effects
$\mathcal{K}=\mathcal{K}_0 \left(1+\sum_i \frac{\alpha_i
\hbar^i}{(r_+^2+a^2)^i}\right)^{-1}$, where $\alpha_i$ correspond to
higher order loop corrections to the surface gravity of black holes
and $\mathcal{K} = 2 \pi T$. Thus the entropy including the
correction terms becomes

\begin{eqnarray}\label{soln6}
S &=& \frac{\pi}{\hbar}(r_+^2+a^2)+ \pi \alpha_1 ln (r_+^2+a^2)+
\sum_{k>2} \frac{\pi \alpha_{k-1}
\hbar^{k-2}}{(2-k)(r_+^2+a^2)^{k-2}} + \cdots.
\end{eqnarray}

Note that we can get the corrections for the Kerr black black hole
\cite{solod1} if we put charge $Q=0$, the Schwarzschild black hole,
$a=Q=0$ and the Reissner-Nordstr\"{o}m black hole, $a=0$.

Using the Bekenstein-Hawking area law relating entropy and horizon
area, $S=A/4\hbar$, where the area in our case is $A=4\pi
(r_+^2+a^2)$, from (\ref{soln6}) we obtain the modified area law as

\begin{eqnarray}\label{soln7}
S &=& \frac{A}{4 \hbar}+ \pi \alpha_1 ln A -\frac{4 \pi^2 \alpha_2
\hbar}{A}-\frac{8 \pi^3 \alpha_3 \hbar^2}{A^2}-\cdots .
\end{eqnarray}

%\section{Axially symmetric Einstein-Maxwell dilaton-axion bh}

Now, we consider the stationary axisymmetric Einstein-Maxwell black
holes in the presence of dilaton-axion  field, found in heterotic
string theory \cite{strom}. In Boyer-Lindquist coordinates these are
described by \cite{jialing}

\begin{eqnarray}\nonumber
ds^{2} &=& - \frac{\Sigma - a^{2}sin^{2}\theta}{\Delta} dt^{2} -
\frac{2a sin^{2}\theta}{\Delta}\left[(r^{2}
-2Dr+a^{2})-\Sigma\right]dt d\phi\\ &+& \frac{\Delta}{\Sigma} dr^{2}
+ \Delta d\theta^{2}  + \frac{sin^{2}\theta}{\Delta}
\left[(r^{2}-2Dr+a^{2})^2-\Sigma a^{2}sin^{2}\theta\right]d\phi^{2}
,
\end{eqnarray}
where $\Delta = r^{2} -2Dr + a^{2}cos^{2}\theta ,  \Sigma =
r^{2}-2Mr + a^{2}$.

They have the electric charge $Q = \sqrt{2\omega D(D-M)}$, where
$\omega = e^{d}$. Here $D$, and $d$ denote the dilaton charge and
the massless dilaton field, and $m = M-D$ is the
Arnowitt-Deser-Misner (ADM) mass of the black hole. The
electrostatic potential is $\Phi = (-2DM/Q(r_{+}^{2} -2Dr_{+}+
a^{2})$. The angular velocity on the horizon is given by

\begin{equation}\nonumber
 \Omega = \frac{J}{2M\left[M(M+D)+\sqrt{M^2(M+D)^2-J^2}\right]}
\end{equation}

The metric has singularities at $r^{2} -2Dr+ a^{2}cos^{2}\theta =
0$. The outer and inner horizons are respectively $r_{\pm} =
\left(M-Q^2/2\omega M \right) \pm \sqrt{\left(M-Q^2/2\omega M
\right)-a^2}$.

The Hawking temperature is \cite{MAKS-EPJC}
\begin{equation}\label{tempda}
T = \frac{\hbar}{4\pi }\left[\frac{
\sqrt{M^2(M+D)^2-J^2}}{M[M(M+D)+\sqrt{M^2(M+D)^2-J^2]}}\right] .
\end{equation}

One can easily check that the above thermodynamical quantities
satisfy conditions (\ref{3conda})-(\ref{3condc}). Thus the entropy
differential $dS$ is exact and we can work out the entropy
corrections as

\begin{eqnarray*}\label{soln60}
S &=& \frac{\pi}{\hbar}(r_+^2-2Dr_{+}+a^2)+ \pi \beta_1 \ln
(r_+^2-2Dr_{+}+a^2) \\ &+& \sum_{k>2} \frac{\pi \beta_{k-1}
\hbar^{k-2}}{(2-k)(r_+^2-2Dr_{+}+a^2)^{k-2}} + \cdots .
\end{eqnarray*}

The Bekenstein-Hawking entropy associated with this horizon is one
quarter of the area of the horizon surface. It is important to note
that unlike spherical geometry the horizon surface here is not
simply a 2-sphere. The area of the horizon from the 2-metric on the
horizon is $A = 4\pi (r_{+}^{2} -2Dr_{+} + a^{2})$, and using the
corresponding entropy $S = \frac{A}{4\hbar}$, we again obtain the
area law given by Eq. (\ref{soln7}).

\acknowledgments

KS acknowledges travel grant from the Higher Education Commission of
Pakistan to participate in MG12 in July 2009, at UNESCO, Paris.

\end{document}